# Multi-command Tactile and Auditory Brain Computer Interface based on Head Position Stimulation

H. Mori[1], Y. Matsumoto[1], Z.R. Struzik[2,3], K. Mori[4], S. Makino[1], D. Mandic[5], and T.M. Rutkowski[1,2]

[1]Life Science Center of TARA, University of Tsukuba, Japan; [2]RIKEN Brain Science Institute, Wako-shi, Japan; [3]The University of Tokyo, Tokyo, Japan; [4]Research Institute of National Rehabilitation Center for Persons with Disabilities, Tokorozawa, Japan; [5]Imperial College London, London, UK

Correspondence: T.M. Rutkowski, Life Science Center of TARA, University of Tsukuba, 1-1-1 Tennodai, Tsukuba, Ibaraki, Japan.
E-mail: tomek@tara.tsukuba.ac.jp

*Abstract.* We study the extent to which vibrotactile stimuli delivered to the head of a subject can serve as a platform for a brain computer interface (BCI) paradigm. Six head positions are used to evoke combined somatosensory and auditory (via the bone conduction effect) brain responses, in order to define a multimodal tactile and auditory brain computer interface (taBCI). Experimental results of subjects performing online taBCI, using stimuli with a moderately fast inter-stimulus interval (ISI), validate the taBCI paradigm, while the feasibility of the concept is illuminated through information transfer rate case studies.

## 1. Introduction

State of the art BCI relies mostly on mental visual and motor imagery paradigms, which require long training and non-impaired vision of the subjects. Recently, alternative solutions have been proposed to utilize spatial auditory [Halder et al., 2010; Schreuder et al., 2010] or tactile (somatosensory) modalities [Muller-Putz et al., 2006; van der Waal et al., 2012] in order to enhance brain-computer interface comfort, or to boost the information transfer rate (ITR) achieved by users. The concept described in this paper, of utilizing brain somatosensory (tactile) modality, opens up the attractive possibility of targeting the tactile sensory domain, which is not as demanding as vision during the operation of robotic interfaces (wheelchair, prosthetic arm, etc.) or visual computer applications. The first successful trial to utilize somatosensory modality to create a BCI [Muller-Putz et al., 2006] targeted a very low stimulus frequency in a range of 20-31 Hz to elucidate the subject's attentional modulation of steady-state responses. A very recent report [van der Waal et al., 2012] proposed using a Braille stimulator with a 100 ms long static push stimuli delivered to six fingers to evoke somatosensory response related P300. Very encouraging results were obtained with 7.8 bit/min on average and 27 bit/min for the best subject. Here we propose to combine the two above-mentioned modalities in a taBCI paradigm, which relies on a P300 response evoked by the audio and tactile stimuli delivered simultaneously via the vibrotactile exciters attached to positions on the head, thus benefiting from the bone-conduction effect for audio, which could help ALS-TLS (Amyotrophic Lateral Sclerosis / Total Locked-in State) patients with compromised vision and hearing due to weakened blinking and middle ear effusion/negative pressure, respectively. This offers a viable alternative for individuals lacking somatosensory responses from the fingers.

## 2. Material and Methods

In the experiments reported in this paper, eleven BCI-naïve subjects took part (mean age 21.82 years, with a standard deviation of 0.87). All the experiments were performed at the Life Science Center of TARA, University of Tsukuba, Japan. The psychophysical and online EEG taBCI paradigm experiments were conducted in accordance with the WMA Declaration of Helsinki - Ethical Principles for Medical Research Involving Human Subjects. The subjects of the experiments received a monetary gratification. The 100 ms long stimuli in the form of 350 Hz sinusoidal waves were delivered to areas of the subjects' heads via the tactile exciters HiWave HIAX19C01-8 working in the range of 300-20,000 Hz. The vibrotactile stimulators were arranged as follows. The pairs of exciters were attached on both sides of the forehead, chin, and behind the ears respectively. During the online taBCI experiments the EEG signals were captured with an eight dry electrodes portable wireless EEG amplifier system, g.MOBllab+ & g.SAHARA by g.tec. The electrodes were attached to the following head locations Cz, CPz, P3, P4, C3, C4, CP5, and CP6, as in the 10/10 extended international system (see topographic plot in Fig. 1). The ground and reference electrodes were attached behind the left and right ears respectively. In order to limit electromagnetic interference, the subjects' hands were additionally grounded with armbands connected to the amplifier ground. No electromagnetic interference was observed with the vibrotactile exciters attached to the head. The recorded EEG signals were processed by an in-house enhanced BCI2000 application using a linear discrimination analysis (LDA) classifier with features drawn from 0-600 ms event related potential (ERP) intervals. The sampling rate was set to 256 Hz, high pass filter at 0.1 Hz, and low pass filter at 40 Hz. The ISI was 400 ms and each stimuli length was 100 ms. The subjects were instructed to spell six-digit random sequences of numbers 1-6, which were represented by the exciters in each session. Each target was presented five times in a single spelling trial, and the averages of five ERPs were later used for the classification.

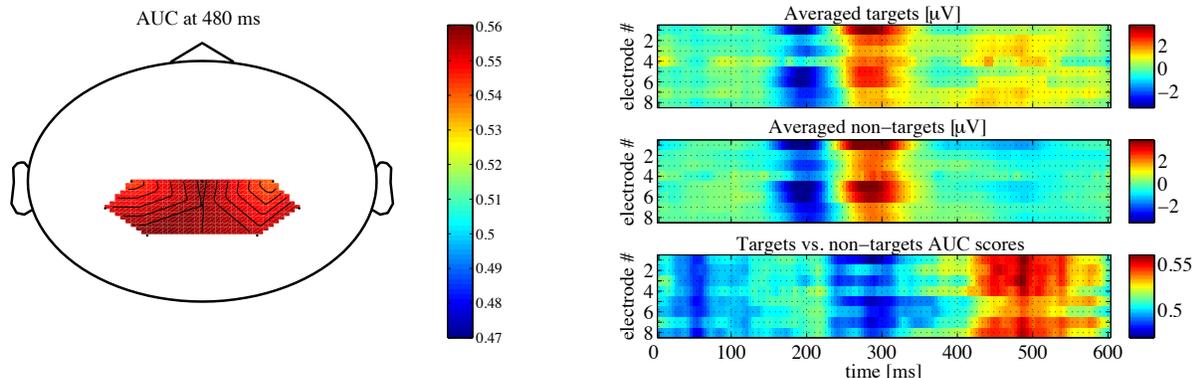

*Figure 1.* The averaged ERP responses of the eleven BCI-naïve subjects taking part in the experiment. The left panel presents a topographical plot with the target vs. non-target ERP area under the curve (AUC) for ERP discriminative features extraction [Schreuder et al., 2010] plotted at the maximum value at 480 ms. The right panel presents the averaged ERPs (11 subjects; averaged six-digit sequences spelled in three sessions by each subject) for targets in the top panel; non-targets in the middle; and AUC coefficients at the bottom. The EEG electrode, represented by numbers on the vertical axes is shown in the right panels. The electrode order is as follows Cz, CPz, P3, P4, C3, C4, CP5, CP6.

## 3. Results and Discussion

The averaged ERP responses from the eleven BCI-naïve subjects are presented in Fig. 1 for the target and non-target digits separately, together with the area under the curve (AUC) [Schreuder et al., 2010] discrimination coefficient plots marking the most separable latencies. A topographic plot of the AUC coefficient distributions is also presented in Fig. 1, supporting the choice of the eight dry EEG electrodes covering the parietal cortex. The results of online BCI interfacing sessions are summarized in Table 1 in the form of mean accuracies above the theoretical chance level of 16.6%. In our experiments, only one BCI-naïve subject obtained 100%, and likewise one obtained 0% for the six-digit sequence spelling accuracy with the 5-trials averaging procedure. The preliminary, yet encouraging results presented are a step forward in the search for new BCI paradigms for ALS-TLS patients with compromised vision and hearing symptoms.

*Table 1.* Summary of the online taBCI interfacing results with the 11 naïve subjects (one subject scored 100% accuracy reaching 12.9 bit/min and one 0% with 0 bit/min).

| ERP averages | Mean accuracy | Accuracy standard deviation | Mean ITR | ITR standard deviation |
|---|---|---|---|---|
| 5 | 50% | 27% | 3.22 bit/min | 3.6 bit/min |

**Acknowledgements**

This research was supported in part by the Strategic Information and Communications R&D Promotion Programme no. 121803027 of The Ministry of Internal Affairs and Communication in Japan, and by KAKENHI, the Japan Society for the Promotion of Science grant no. 12010738. We also acknowledge the technical support of YAMAHA Sound & IT Development Division in Hamamatsu, Japan.